# Ultra-Fast Ising Machine for Combinatorial Optimization using Sub-nanosecond CMOS-integrated Magnetoresistive Random Access Memory

**Authors:** Sai Li[1,2,†], Yihao Zhang[1,2,†], Albert Lee[3], Zheng Zhu[3], Lang Zeng[1,2], Peng Wang[1,2], Lei Gao[4], Di Wu[3], Weisheng Zhao[1,2,*]

[1]School of Integrated Circuit Science and Engineering, Beihang University, Beijing, 100191, China.

[2]National Key Lab of Spintronics, Institute of International Innovation, Beihang University, Hangzhou, 311115, China.

[3]InstonTech, Suzhou, 215000, China.

[4]Empyrean Technology Co., Ltd., Beijing, 100102, China.

[†]These authors contributed equally: Sai Li, Yihao Zhang

[*]Corresponding author. Email: weisheng.zhao@buaa.edu.cn

## Abstract

Physics-inspired Ising machines are emerging as promising hardware alternatives to traditional von Neumann architectures for tackling computationally intensive combinatorial optimization problems (COPs). However, the practical application of existing quantum, optical, and electronic platforms is fundamentally constrained in speed and scalability by the physical mechanisms governing their spin dynamics. Here, we report, to our knowledge, the first complementary metal-oxide-semiconductor-integrated chip-scale spintronic Ising machine that enables sub-nanosecond probabilistic spin updates. Exploiting the voltage-controlled magnetic anisotropy effect in dense magnetoresistive random access memory, our design eliminates large write currents and achieves a tunable 0-100% single-pulse switching probability via pulse-width control. Compared with prior spintronic implementations, it achieves a simultaneous 100× improvement in spin-update speed (0.3-1 ns) and energy efficiency (< 40 fJ). Leveraging the all-to-all Ising machine implemented on-chip, we validate its generality on industry-relevant COPs in the electronic design automation domain, including global routing and layer assignment, attaining high-quality solutions with a system energy efficiency of $2.5\times10^4$ solutions per second per watt. This performance outperforms state-of-the-art quantum and graphics processing units by six and seven orders of magnitude on identical benchmarks, respectively. Our results open the sub-nanosecond regime of probabilistic spin updates in a chip-scale Ising machine, establishing spintronics as a compelling route for ultra-fast and scalable physics-inspired intelligence.

## Introduction

Non-deterministic polynomial-time (NP)-hard combinatorial optimization problems (COPs) inherently challenge conventional von Neumann architectures, especially in energy-constrained and compute-intensive applications. The exponentially increasing resource demands with problem size[1] critically limit advancements across diverse scientific and industrial domains, such as artificial intelligence[2], materials science[3], finance[4], transportation[5], communications[6], and integrated circuit design[7]. Physics-inspired Ising machines have emerged as a promising alternative for tackling COPs over the past decade[8]. By mapping these problems onto the Ising model[9], such machines harness Ising spin dynamics to efficiently converge toward ground states, demonstrating remarkable efficiency for finding high-quality solutions[10,11].

A variety of physical implementations have been demonstrated for Ising machines. Quantum processing units (QPUs)[12,13], while powerful in principle[14], contend with cryogenic operating temperatures and sparse topological constraints[15]. Optical approaches, exemplified by coherent and photonic Ising machines, are often constrained by the difficulty of simultaneously ensuring full programmability and scalability[16–20]. Conventional complementary metal-oxide-semiconductor (CMOS) approaches, spanning graphics processing units (GPUs)[21], field-programmable gate arrays (FPGAs)[22], and application-specific integrated circuits[23–28], are widely explored but often suffer from significant digital overheads for pseudo-random number generators (PRNGs)[29,30] and computational latency due to ex-situ Ising spin updates[31].

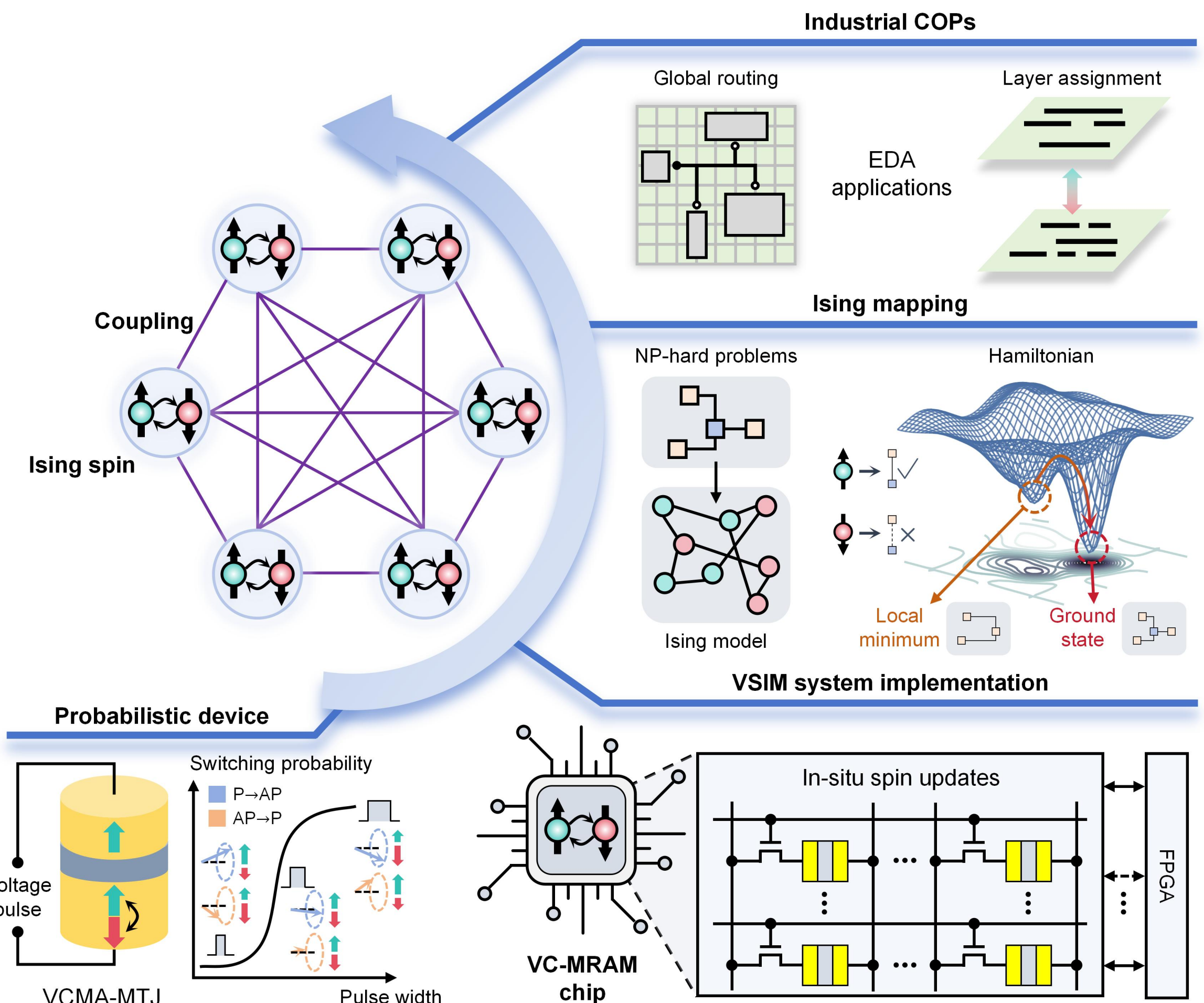

**Fig. 1 | Overview of the voltage-controlled spintronic Ising machine (VSIM).** (i) Probabilistic device: Utilization of intrinsic pulse-width-tunable probabilistic switching in VCMA-MTJs for in-situ Ising spin updates. (ii) Hardware implementation: Realization of a chip-scale Ising machine enabled by the CMOS-compatible fabrication of VC-MRAM. (iii) Ising mapping: Translation of real-world NP-hard COPs onto the Ising Hamiltonian, with the global energy minimum representing the optimal solution. (iv) COP applications: Validation of the system capability through global routing and layer assignment tasks within the EDA domain.

To overcome these limitations, recent research has focused on CMOS-compatible nanodevice-based Ising spins, exploiting distinctive physical properties to perform efficient computing at scale. Implementations reported to date span nano-oscillators[32–36] and stochastic devices[37–43]. Spintronic devices—particularly magnetic tunnel junctions (MTJs)—represent the most promising candidates due to their intrinsic stochasticity and high integration density. Prior works have primarily centered on current-driven MTJs, including superparamagnetic tunnel junctions (SMTJs)[37–40] and spin-orbit torque (SOT) MTJs[41], which provide the necessary tunable probability through spin-torque-driven dynamics. Most recently, voltage-controlled magnetic anisotropy (VCMA)-MTJs[43] use multiple devices as random number generators, combined with post-processing circuits, to emulate Ising spins. Crucially, these approaches still require CMOS-integrated chip-scale validation to fully confirm their potential for high-speed and energy-efficient operation (surpassing current benchmarks of ~100 ns and ~1 pJ), as well as a definitive advantage in tackling industrial COPs.

In this work, we demonstrate the first chip-scale spintronic Ising machine using 96-kb voltage-controlled MRAM (VC-MRAM). By leveraging the VCMA-MTJ, which eliminates large write currents and their associated spin-torque-driven delays, we achieve in-situ Ising spins by tuning write pulse widths, enabling probabilistic spin updates at an ultra-fast speed (< 1 ns) with an energy cost below 40 fJ. This represents a simultaneous 100× improvement in both speed and energy efficiency over conventional spintronic schemes. We experimentally solve two representative NP-hard COPs from very large-scale integration (VLSI) electronic design automation (EDA)—global routing and layer assignment—using our Ising machine. It achieves $2.5\times10^4$ solutions per second per watt, surpassing state-of-the-art Ising machines by three to seven orders of magnitude. These findings highlight spintronics as a high-speed, energy-efficient route to physics-inspired computing, enabling Ising machines to solve complex industrial-scale optimization challenges.

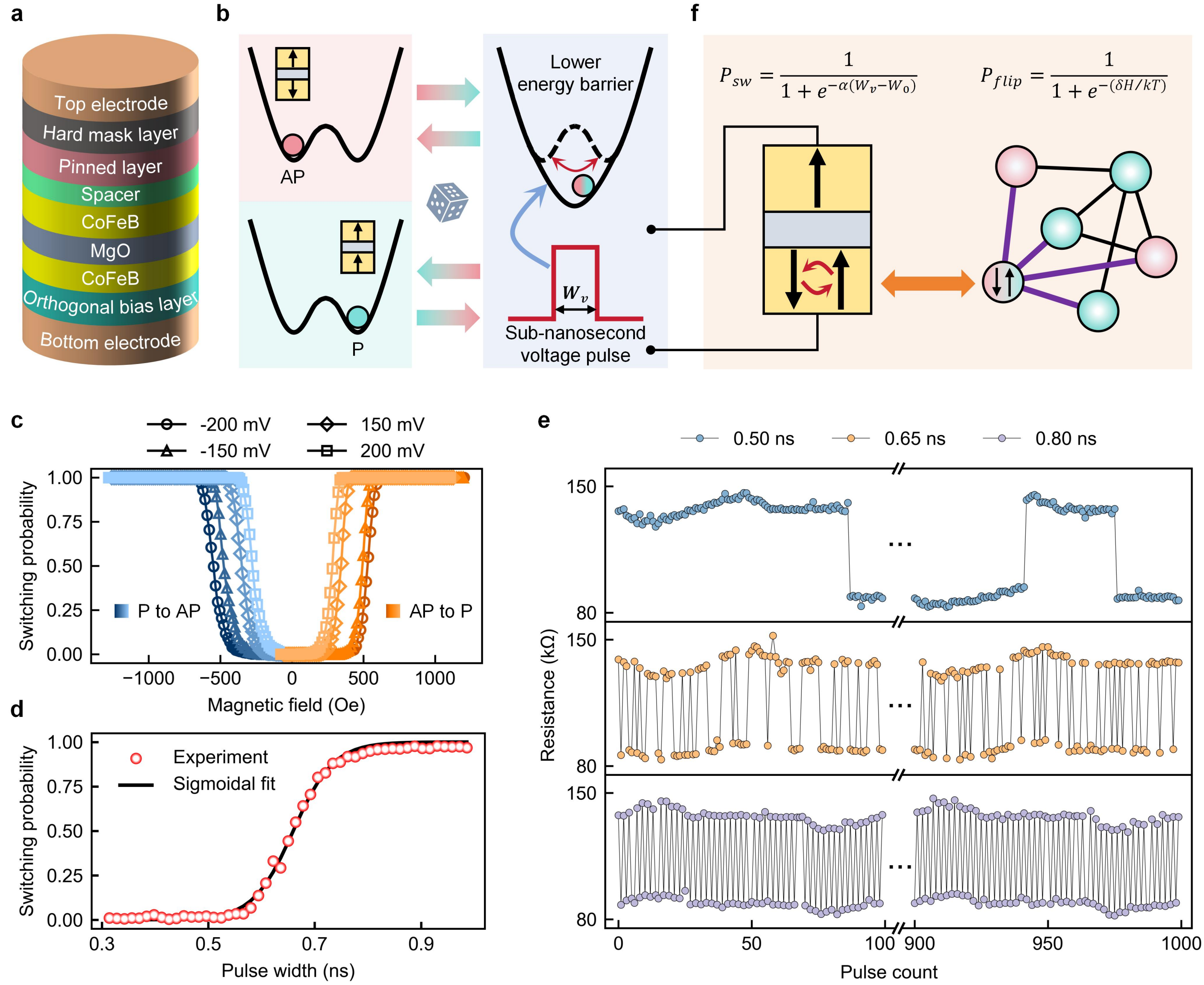

**Fig. 2 | Implementation of Ising spins using our VCMA-MTJ with tunable switching probability. a,** VCMA-MTJ multilayer stack architecture, where the free-layer magnetization is controlled by a voltage across the MgO barrier, while the reference layer remains fixed. **b,** The VCMA effect reduces the energy barrier, enabling probabilistic switching between P and AP states. **c,** $P_{sw}$ versus magnetic field under different voltage biases. It indicates the symmetry of the P-to-AP and AP-to-P switching. **d,** Sigmoidal $P_{sw}$ versus $W_v$ (sub-nanosecond) under a write voltage of 2.2 V. **e,** Resistance measurements after each of 1000 consecutive 0.5 ns, 0.65 ns, and 0.8 ns write pulses, corresponding to $P_{sw}$ of ~2%, ~45% and ~95%. **f,** Voltage-controlled probabilistic switching of the VCMA-MTJ enables in-situ ultra-fast Ising spin updates.

## Sub-nanosecond Probabilistic MTJ for Ising Spins

In an Ising machine, spins interact and evolve dynamically toward the ground state, providing an effective framework for solving COPs[8], where solutions are obtained by minimizing the Ising Hamiltonian. For a fully connected Ising machine with all-to-all spin interactions, the Hamiltonian is expressed as

$$H = -\sum_{i<j} J_{i,j} s_i s_j - \sum_i h_i s_i \tag{1}$$

where $s_i$ denotes the $i$-th spin that can adopt either of two states $\{-1, +1\}$, corresponding to spin-down or spin-up configurations. The coupling strength between spins is characterized by the matrix $J$, and $h$ represents the external magnetic field. $J$ and $h$ define the system's energy landscape, whereas spin dynamics are essential for the efficient exploration of it. Such dynamics can be modeled using the well-established Glauber approach[44]:

$$P_{flip} = \frac{1}{1+e^{-(\delta H/T)}} \tag{2}$$

where $P_{flip}$ denotes the probability of spin flipping, $\delta H$ represents the energy reduction in the system Hamiltonian upon flipping, and $T$ is the effective temperature. To facilitate fast convergence toward low-energy states, an annealing process is typically employed, during which $T$ is gradually decreased in an iterative manner. The probabilistic flipping of Ising spins, representing thermal fluctuations in physical systems, is crucial for escaping local minima, as it allows flips even when $\delta H$ is negative (indicating an energy increase), provided $T$ is sufficiently high.

An essential requirement for implementing an Ising machine is an efficient building block capable of realizing Ising spins with a tunable $P_{flip}$. Most conventional approaches[37–42,45,46] employ current-driven MTJs as Ising spins, where electrical current induces spin torques to probabilistically switch the magnetization between the anti-parallel (AP) and parallel (P) states (representing the $+1$ and $-1$ states of Ising spins, respectively). In such MTJ devices, magnetization states still exhibit long retention times (μs to ms) and suffers from high energy consumption for switching. To address these limitations, we adopt perpendicular VCMA-MTJ devices (Fig. 2a) as Ising spins, where the bistable magnetization switching probability ($P_{sw}$) is purely controlled by an electric field. As shown in Fig. 2b, the VCMA effect modulates the interfacial magnetic anisotropy and thereby reduces the energy barrier between AP and P states. It leads to the probabilistic magnetization precession[47] at sub-nanosecond scale, with $P_{sw}$ dependent on the width of the single voltage pulse. The MTJ is designed with a high resistance-area product (RA) of 300 Ω·μm$^2$, where a thick MgO layer serves to eliminate spin-transfer torque current contribution and reduce the switching energy consumption significantly. It is worth mentioning that the VCMA-MTJ achieves a symmetry of the switching process as demonstrated in Fig. 2c, where the $P_{sw}$ values for AP-to-P and P-to-AP transitions are nearly identical under equivalent voltage conditions. As a result, the probabilistic flipping required for an Ising spin can be induced in the VCMA-MTJ by the same voltage pulse irrespective of the initial P or AP state, obviating the need for reset operations of the device. To accurately evaluate the $P_{sw}$, we perform 1000 repeated measurements of the MTJ state, each following the application of an identical voltage pulse. The voltage pulse width ($W_v$) is modulated within the range of 0.3-1 ns to obtain the corresponding $P_{sw}$ values. As shown in Fig. 2d, $P_{sw}$ as a function of $W_v$ exhibits a sigmoid dependence:

$$P_{sw} = \frac{1}{1+e^{-\alpha(W_v-W_0)}} \tag{3}$$

where $\alpha$ and $W_0$ represent fitting coefficients. Fig. 2e illustrates the resistance measurements obtained after individual pulses with durations of 0.5 ns, 0.65 ns, and 0.8 ns, corresponding to $P_{sw}$ of approximately 2%, 45%, and 95%, respectively. These results demonstrate that the VCMA-MTJ devices can span the full $P_{sw}$ range from 0-100% within a voltage amplitude range of 1.9-2.2 V (Supplementary Note 1). To ensure the in-situ annealing process, the $P_{sw}$ of VCMA-MTJs must match the $P_{flip}$ of Ising spins. In this framework, the applied pulse width for voltage-controlled probabilistic switching is given by

$$W_v = \frac{\delta H}{\alpha T} + W_0 \tag{4}$$

This sigmoidal switching characteristic, controlled by sub-nanosecond voltage pulses, renders the VCMA-MTJ a direct physical embodiment of Ising spins (Fig. 2f).

The advantages of modulating the pulse width of the VCMA-MTJ are as follows. (1) The write time of the Ising spin device can be reduced to < 1 ns, due to the intrinsic property of the VCMA effect to modulate the energy barrier directly. However, current-driven MTJs rely on spin torques to induce stochastic magnetization switching. Their narrow operating window (~ 0.1 V) imposes significant constraints on integrated circuit design, making them less suitable for stable Ising chips requiring fine-grained probability resolution. It is also worth noting that compared to current-driven implementations, the VCMA-MTJ ensures enhanced device endurance because frequent Ising spin updates do not require current injection through the MTJ stack. Our experimental results confirm exceptional device endurance up to $10^{13}$ cycles (see Supplementary Data), with further testing underway to assess longevity beyond this limit. (2) By simply adjusting the pulse width of our VCMA-MTJ, probability values spanning 0-100 % can be readily implemented, offering both low latency and compact circuit area. Other approaches based on 50 %/50 % stochastic bit streams require multiple bits to generate a uniform random number for spin updates. For example, producing a 16-bit random number with a single MTJ would require 16 successive cycles. Moreover, such methods depend on large-area peripheral circuitry—often including nonlinear function units and comparators—to achieve sigmoidal probability sampling, resulting in significant area and complexity overhead. Overall, based on the proposed superior VCMA-MTJ Ising spin, an Ising chip has been developed to implement VSIM, as detailed in the next section.

## Hardware implementation of VSIM

While prior spintronic Ising machines[37,39,41,43] demonstrated the concept using discrete MTJs, they have not yet exploited the full potential of in-situ spins at the chip scale. Here, we bridge this gap by fabricating a 96-kb VC-MRAM chip (Fig. 3a) using a 40-nm CMOS process. The integrated VCMA-MTJs (~70 nm diameter) exhibit negligible die-to-die resistance variations (Supplementary Fig. 1e), ensuring uniform switching probabilities. And the high-quality randomness of our VC-MRAM is confirmed via the National Institute of Standards and Technology tests (Supplementary Note 2).

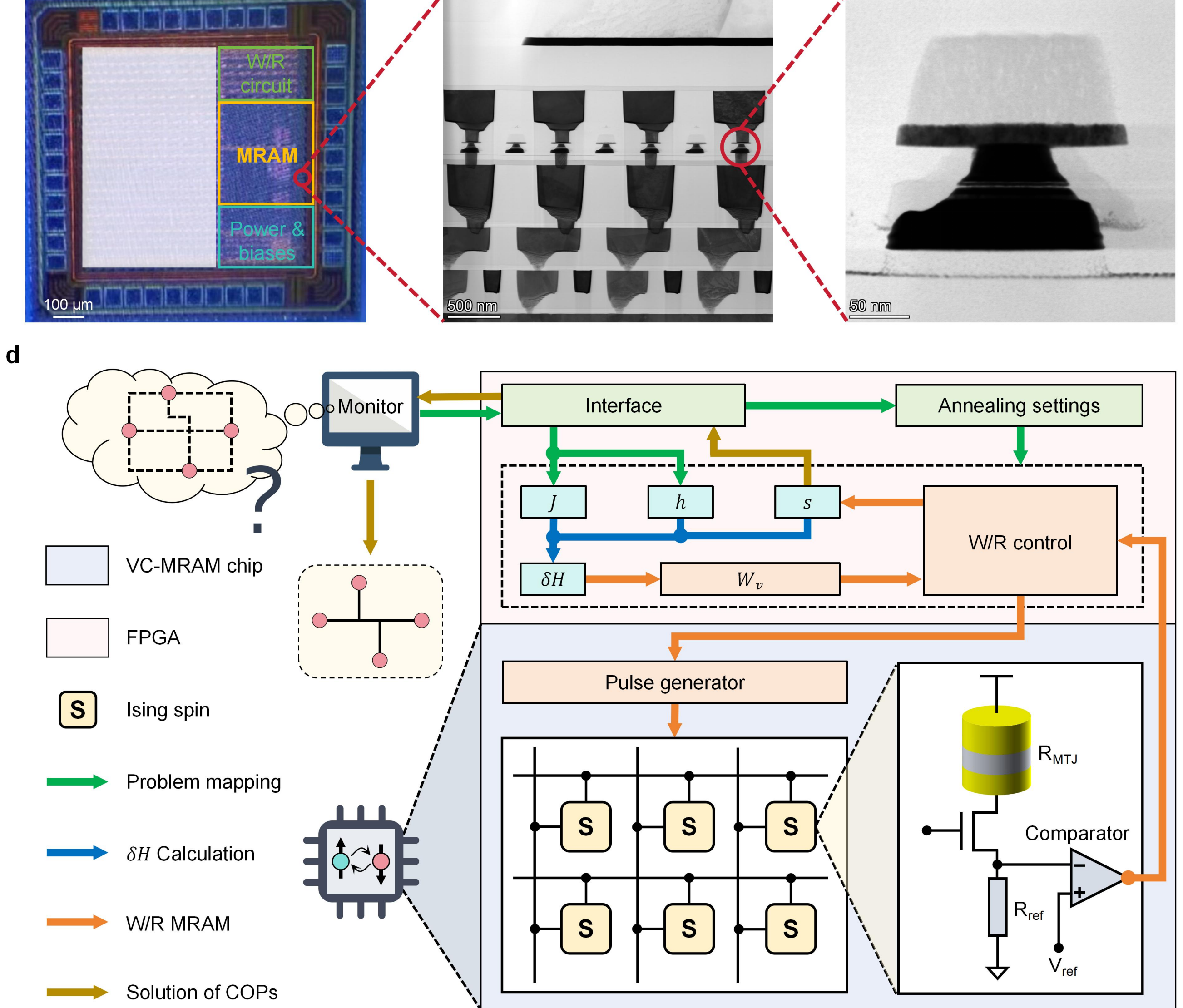

**Fig. 3 | Implementation of VSIM with the VC-MRAM chip. a,** Die micrograph of the fabricated VC-MRAM chip. **b,** Cross-sectional TEM image illustrating the vertical integration of VCMA-MTJs with CMOS circuitry. **c,** TEM image of a single VCMA-MTJ with a nominal diameter of 70 nm. **d,** VSIM architecture: (i) The host monitor interfaces with the FPGA to configure Ising parameters $J$ and $h$; (ii) The FPGA modulates the on-chip pulse generator to drive the VCMA-MTJs for in-situ spin updates; (iii) Spin states $s$ are sampled for subsequent computation.

To enable in-situ Ising spin updates, an integrated pulse generator within the VC-MRAM (see Supplementary Note 3) applies precisely timed voltage pulses to the VCMA-MTJs. The VSIM is implemented on our VC-MRAM chip, with system-level control and data interfacing orchestrated by an FPGA. The system design diagram is illustrated in Fig. 3d, and the experimental prototype is shown in Extended Data Fig. 1. NP-hard COPs are mapped onto the Ising model by determining the necessary Ising parameters $J$ and $h$ (the specific mapping process will be discussed later). These parameters, along with the VSIM annealing settings, are configured via the monitor. During the annealing process, the energy reduction $\delta H$ of flipping $s_i$ is calculated as

$$\delta H =- 2s_i(\textstyle\sum_j J_{i,j}s_j + h_i) \tag{5}$$

which is subsequently used to derive the corresponding pulse width $W_v$ according to Eq. (4). Based on this $W_v$, the write/read (W/R) control module drives the VC-MRAM's integrated pulse generator. The Ising spin $s_i$ is updated according to the target probability, induced by the precise $W_v$. Upon completion of the annealing iterations, the final spin states representing the solution to the COP are read out to the monitor.

The VC-MRAM chip serves as the core enabler of a highly efficient and general-purpose Ising machine. On the one hand, unlike discrete devices limited by wire-bonded interconnects, our chip ensures the W/R bandwidth critical for a full-stack, ultra-fast Ising machine. Ideally avoiding the latency bottlenecks of discrete configurations, which can lag behind spin dynamics by orders of magnitude[43], we solve this by employing a CMOS-integrated VC-MRAM with high-speed W/R circuits. On the other hand, the VC-MRAM chip delivers improved MTJ stability with reduced access overhead, directly enhancing VSIM programmability. While the current system leverages an FPGA to tackle various NP-hard COPs with flexibility, the underlying design principles readily support a transition to a fully integrated system-on-chip. Regarding density, the VC-MRAM stands out by eliminating the requirement for large write currents, which enables the integration of minimum-sized access transistors. Such scalability is indispensable for handling industrial-scale workloads, including the VLSI EDA applications discussed next.

## VSIM for global routing

VLSI EDA is a powerful tool for the design of increasingly complex integrated circuits in the modern semiconductor industry, a field where many core VLSI design problems are NP-hard COPs[48,49]. The VLSI design process is typically segmented into three critical phases: component placement, physical routing determination, and layer assignment[50]. Here, we start from mapping the global routing problem onto the Ising model as it plays a crucial role in minimizing routing distance across the chip, which is efficiently implemented on-chip using the VSIM system.

The global routing problem can be effectively modeled as a rectilinear Steiner minimum tree (RSMT) problem. To achieve this mapping, consider a directed weighted graph $G = (V, E)$, where the weight of each edge $(u, v) \in E$ is denoted as $c_{u,v}$. Given a set $U \subseteq V$ of terminal vertices and a root vertex $v_0 \in U$, the Steiner minimum tree is a spanning tree $T = (V_T, E_T)$ that satisfies $U \subseteq V_T \subseteq V$ and $E_T \subseteq E$, while minimizing the total cost. Vertices in the set $W = V \setminus U$ are referred to as Steiner vertices[51]. The RSMT problem is a specific variant of the Steiner minimum tree problem in routing scenarios[52] (Fig. 4a). Directly mapping the RSMT problem to the Ising model is not sufficiently concise; therefore, we first employ a quadratic unconstrained binary optimization formulation as proposed in reference [53]. The variables include edge variables $x^e$ and order variables $x^o$, both of which take values in $\{0, 1\}$. Specifically, when $x^e_{u,v} = 1$, it indicates that edge $(u, v)$ is included in the Steiner tree while $x^e_{u,v} = 0$ indicates it is not included. If $x^o_{u,v} = 1$, it signifies that $u$ is closer to the root vertex than $v$, and 0 otherwise (Fig. 4b). The total Hamiltonian for the RSMT problem is stated as:

$$H_{RSMT}(x) = H_o + \lambda_1 H_{c1} + \lambda_2 H_{c2} + \lambda_3 H_{c3} + \lambda_4 H_{c4} \tag{6}$$

Here, $H_o$ represents the optimization objective, specifically the total cost of all edges in the Steiner tree:

$$H_o = \sum_{\substack{(u,v)\in E,\\ v\neq v_0}} c_{u,v} x^e_{u,v} \tag{7}$$

$H_{c1} \sim H_{c4}$ denote four essential constraints for Steiner tree formation, with corresponding penalty coefficients $\lambda_1 \sim \lambda_4$ that must be carefully selected for obtaining the optimal performance. These constraints serve distinct purposes (Fig. 4c). $H_{c1}$ ensures that each terminal vertex (except the root) has precisely one incoming arc:

$$H_{c1} = \sum_{v\in U\setminus\{v_0\}} \left(1 - \sum_{(u,v)\in E} x^e_{u,v}\right)^2 \tag{8}$$

$H_{c2}$ enforces acyclicity, a fundamental requirement for tree structures. $H_{c3}$ penalizes configurations where Steiner vertices either have multiple incoming arcs or possess outgoing arcs without any incoming ones. $H_{c4}$ addresses the rectilinearity constraint specific to routing scenarios. When two vertices $u$ and $v$ cannot be connected by a straight segment, a positive penalty is imposed. $H_{c2}$ is expressed as a function of both $x^e$ and $x^o$, while other constraints are formulated solely in terms of $x^e$ (detailed mathematical formulations for $H_{c2}$, $H_{c3}$, and $H_{c4}$ are provided in Supplementary Note 4). The spin variables $s$ in the Ising model are derived through a composite transformation of $x^e$ and $x^o$:

$$s = \begin{pmatrix} s^e \\ s^o \end{pmatrix} = 2 * \begin{pmatrix} x^e \\ x^o \end{pmatrix} - 1 \tag{9}$$

where $s^e$ and $s^o$ are corresponding Ising spins for edge and order variables, respectively. Therefore, we successfully construct the mapping from the RSMT problem to the Ising model, yielding $J$ and $h$ by comparing Eq. (1) and Eq. (6). The complete derivation process is also detailed in Supplementary Note 4.

After $J$ and $h$ are mapped and configured in VSIM, the VC-MRAM chip initiates the in-situ annealing process, leveraging the intrinsic probabilistic switching of VCMA-MTJs. As shown in Fig. 4d, while the total Hamiltonian exhibits occasional increases, it follows a distinct downward trend overall, converging to the ground state rapidly within 1000 iterations. Throughout this process, the temperature $T$ decreases exponentially to achieve fast convergence. Simultaneously, Ising spin flips are governed by both the system's energy change ($\delta H$) and thermal fluctuations related to $T$, as described in Eq. (3). The flip rate, defined as the proportion of spins that flip in the entire system per iteration during the annealing process, is a statistically measured value. At high $T$, the flip rate approaches 0.5, reflecting a randomized state, while at low $T$, it tends towards 0 as the system stabilizes. Fig. 4e shows the routing results at four representative iterations, illustrating the evolution from a random initial state through suboptimal solutions to the optimal solution. This progression underscores VC-MRAM's ability to provide sufficient stochasticity, enabling the system to escape local minima and converge toward the ground state.

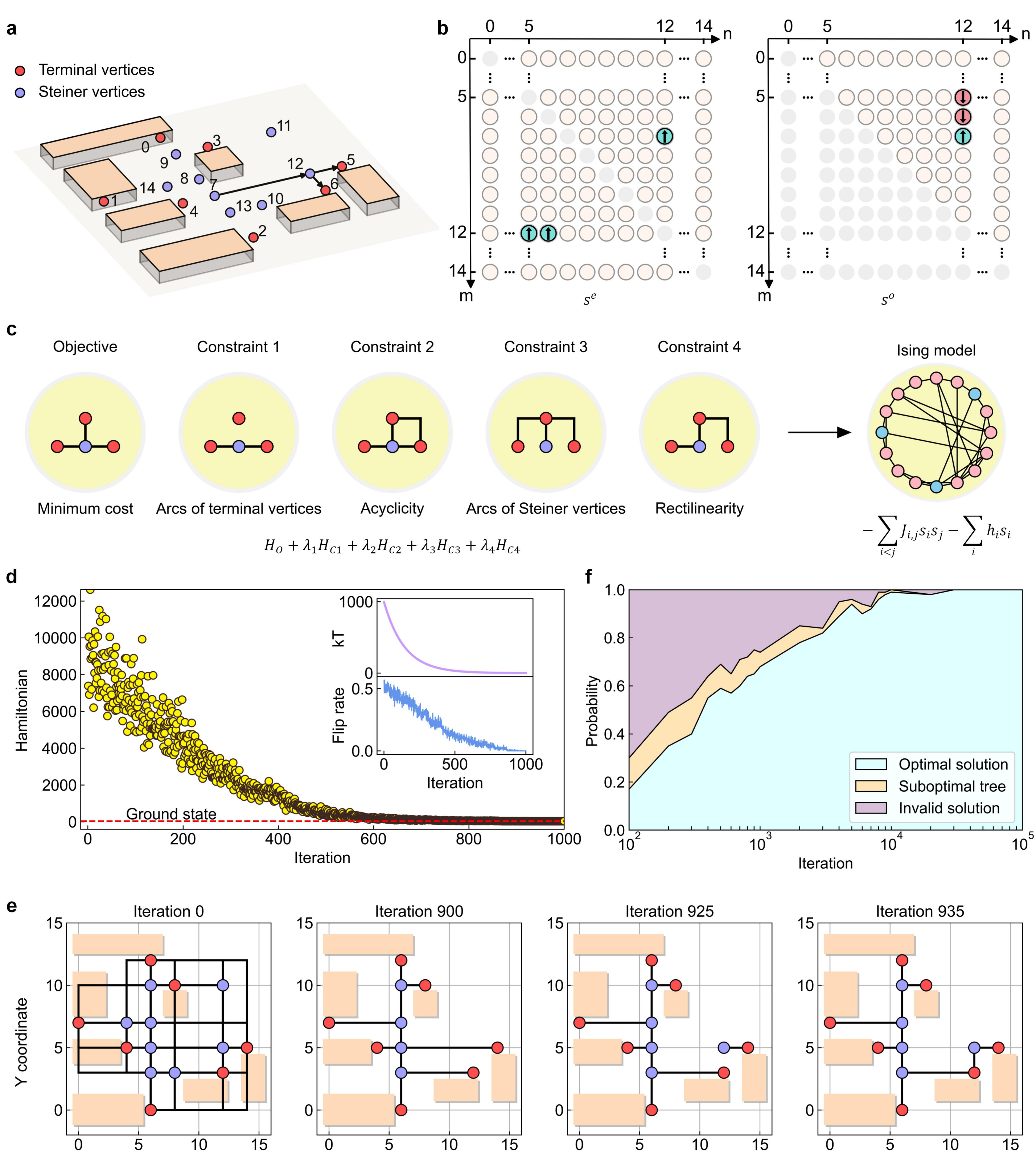

**Fig. 4 | Global routing demonstration on VSIM. a,** Global routing problem formulated as an RSMT problem, where terminal vertices correspond to circuit module pins and Steiner vertices represent auxiliary routing nodes for wirelength optimization. **b,** Variable definitions: $s^e_{i,j}$ encodes edge connectivity between vertices $i$ and $j$ while $s^o_{i,j}$ specifies topological ordering relative to the routing root. **c,** Ising model mapping framework of the RSMT problem with dual components: (i) Objective term minimizes total wirelength cost; (ii) Constraint terms enforce appropriate arcs of terminal vertices and Steiner vertices, acyclicity and rectilinearity. **d,** Hamiltonian evolution trajectory of the problem in **a** across 1000 iterations. Dashed line represents the ground state. Insets: $kT$ profile and flip rate of Ising spins during the annealing process. **e,** Progressive routing solutions extracted at four representative annealing stages in **d. f,** Solution validity statistics from 100 independent trials per parameter set, classifying results as: optimal solutions, suboptimal trees and invalid solutions.

We categorize solutions into three types: optimal solutions (successful routing with minimized wire length), suboptimal trees

(successful routing without minimized wire length), and invalid solutions (unsuccessful routing, i.e., containing unconnected terminal vertices). For penalty coefficients set to $\lambda_1 = \lambda_2 = 10$ and $\lambda_3 = \lambda_4 = 6$, the probability distribution of these categories over iterations is shown in Fig. 4f. The probability of achieving the optimal solution—termed the success probability—exceeds 95% beyond $10^4$ iterations. The calibration of penalty coefficients is crucial for maintaining high success probability, requiring a careful balance between optimization objectives and constraints. Insufficient coefficient values hinder proper Steiner tree construction (resulting in invalid solutions), while excessive values compromise minimal cost achievement (yielding suboptimal trees). A detailed analysis of success probability variations with different coefficient settings is presented in Supplementary Note 5. Notably, the aforementioned Ising mapping method, along with the VSIM-based solution framework, can be applied to a wide range of routing problems[5,54].

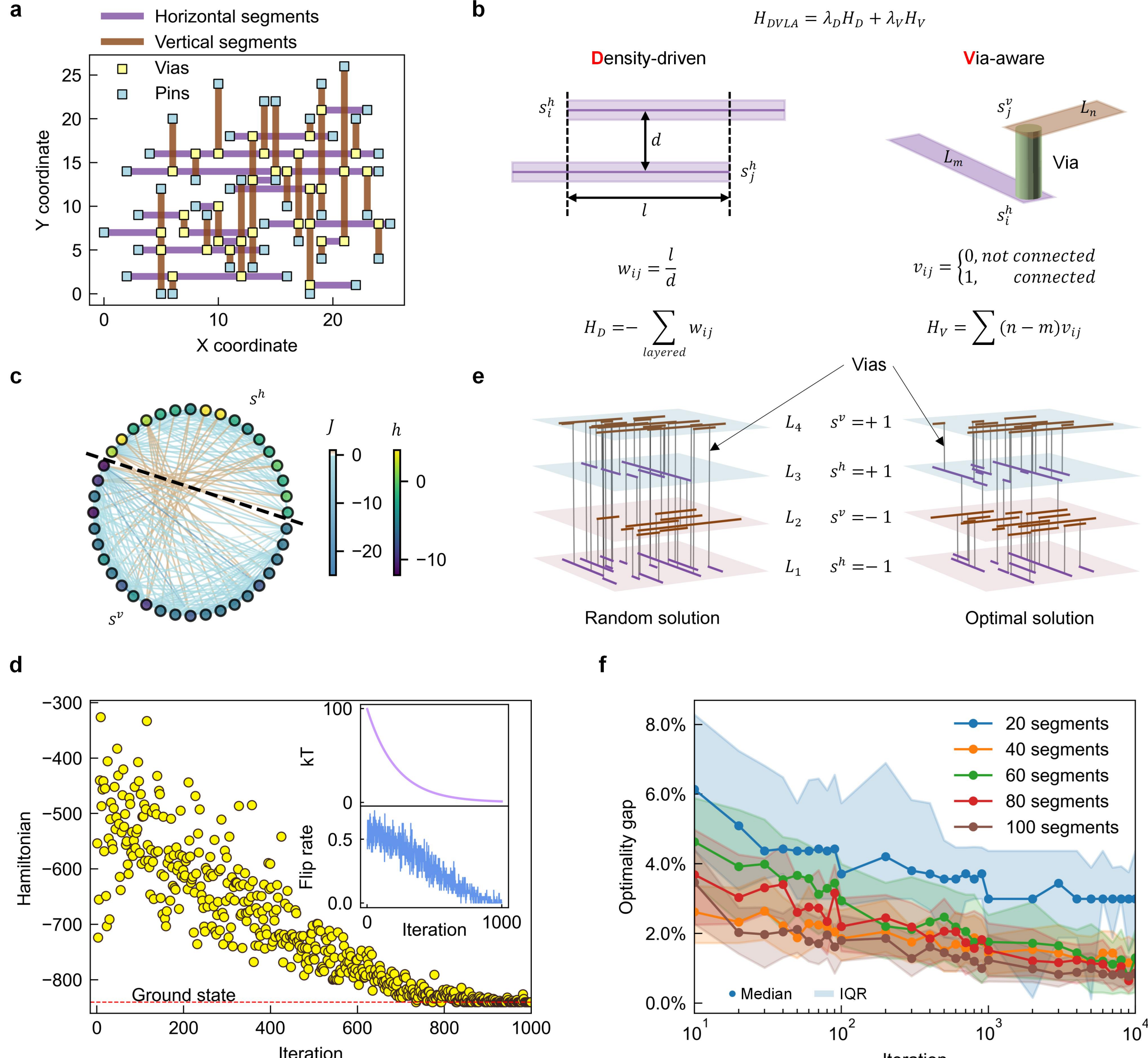

**Fig. 5 | Demonstration of layer assignment optimization using VSIM. a,** Layer assignment scenario illustrating horizontal and vertical wire segments, with inter-layer connections facilitated by vias. **b,** The DVLA problem formulation, which simultaneously optimizes wire density distribution and minimizes via count. **c,** Visualization of the Ising model representation corresponding to the problem in **a.** Circles represent Ising spins, with colors indicating external magnetic field strength. Connection colors between spins denote coupling strength. **d,** Dynamic evolution during problem solving, showing the Hamiltonian trajectory, $kT$ profile, and spin flip rate throughout the annealing process. **e,** Solution visualization demonstrating the dual optimization objectives: homogenized wire density and minimized via count. **f,** Optimality gap versus iteration count for problems of various sizes. Data points represent median values while shaded regions indicate interquartile ranges (IQR).

## VSIM for layer assignment

Layer assignment involves allocating wire segments across multiple metal layers, a task that is considerably more complex than global routing. Following standard practice, we assign horizontal and vertical segments to separate layers to minimize interference. Intersections between horizontal and vertical segments require vias for electrical connectivity (Fig. 5a). This layer assignment strategy addresses two critical issues: managing local wire density[55] and minimizing via usage[50]. Wire density minimization reduces to the Max-cut problem[56] and incorporating via minimization transforms this into a multi-objective optimization challenge. We formulate it as a density-driven via-aware layer assignment (DVLA) problem, mapping it to the Ising model for efficient solution implementation on VSIM.

In our four-layer model ($L_1$-$L_4$), horizontal and vertical segments occupy alternating odd- and even-numbered layers respectively (horizontal in $L_1/L_3$, vertical in $L_2/L_4$). Let Ising spin variables $s = [s^h, s^v]$ directly encode the layer assignment for each segment. If $s_i^h = -1$ ($s_i^v = -1$), it indicates that the $i$-th horizontal (vertical) segment is assigned to $L_1$ ($L_2$); while if $s_i^h = +1$ ($s_i^v = +1$), it is assigned to $L_3$ ($L_4$). The total Hamiltonian for the DVLA problem can be expressed as $H_{DVLA} = \lambda_D H_D + \lambda_V H_V$, where $H_D$ and $H_V$ represent the density-driven and via-aware contributions to the overall objective, respectively. The penalty coefficients $\lambda_D$ and $\lambda_V$ determine the relative weight of each component in the optimization process. Denote the local density matrices of horizontal (vertical) segments as $w^h$ ($w^v$). The local density between two horizontal (vertical) segments is calculated as $l/d$, where $l$ is the overlap length in the horizontal (vertical) direction, and $d$ is the distance in the vertical (horizontal) direction (Fig. 4b). The density-driven term $H_D$ can then be expressed as:

$$H_D = \frac{1}{2}\sum_{0\le i<j<N_h} w_{i,j}^h (s_i^h s_j^h - 1) + \frac{1}{2}\sum_{0\le m<n<N_v} w_{m,n}^v (s_m^v s_n^v - 1) \tag{10}$$

where $N_h$ and $N_v$ denote the numbers of horizontal and vertical segments, respectively. The via-aware term $H_V$ is formulated in Supplementary Note 6, along with a clear derivation of the Ising mapping results. Fig. 5c visualizes the corresponding Ising model of the problem presented in Fig. 5a.

Fig. 5d shows the the rapid decrease of the Hamiltonian, reaching the ground state over 1000 iterations with coefficients $\lambda_D = 10$ and $\lambda_V = 4$. Due to the max-cut nature of $H_D$, the total Hamiltonian becomes negative when $\lambda_D$ is significantly larger than $\lambda_V$. This does not affect the annealing process, as the relative variation of the Hamiltonian, rather than its absolute value, is of primary concern. Fig. 5e illustrates the initial and final solutions to the DVLA problem, demonstrating both homogenized local density and minimized via count. Converging to the optimal solution for the DVLA problem is particularly challenging due to its two competing objectives: reducing wire density by assigning segments to different layers, which inherently increases via requirements, versus minimizing the overall via count. Fig. 5f shows the optimality gap (defined as the relative error between convergence value and minimum value) versus iterations for problems of varying sizes. As iterations exceed $10^4$, the optimality gaps fall below 2%, which is sufficient for real-world EDA designs. Given the capability to explore complex parameter spaces, VSIM is thus well-suited to efficiently tackle such optimization problems characterized by competing objectives and challenging energy landscapes.

## Discussions

To assess the impact of Ising spin quality, we first benchmark our probability-tunable VCMA-MTJs against CMOS implementation using linear feedback shift registers (LFSRs) in solving the global routing problem (Extended Data Fig. 3). We found that VCMA-MTJs achieved a success probability comparable to that of a 16-bit LFSR. Building on this efficiency, VCMA-MTJs demonstrated significant performance advantages over conventional hardware (see Supplementary Notes 7-9): $10^4$ times faster operation with $10^9$ times lower energy consumption than central processing unit (CPU) and GPU; 10× higher speed, 150× lower energy consumption, and 3000× fewer transistors than FPGA. Furthermore, VCMA-MTJs exhibit remarkable resilience to device variations, sustaining > 95% success probability even when a fifth of the devices operate with a peak switching probability (PSP) as low as 60%. Such robustness underscores the viability of Ising machines built on current spintronic chip technologies for tackling inter-device variability, a critical hurdle in emerging devices.

Evaluating overall system performance, we benchmark VSIM against state-of-the-art Ising machines in Table 1, including GPU-based Ising solver[21], QPU[13,14], coherent Ising machine (CIM)[16], phase-transition nano-oscillators (PTNO)[35], SOT-MTJ[41], SMTJ[39] and V-MTJ[43]. A key differentiator is our utilization of VCMA-MTJs, which provide voltage-tunable probabilities to enable ultra-fast (< 1 ns) and low-energy (< 40 fJ) individual spin updates. Benefiting from the VC-MRAM chip, VSIM is a low-power (40 mW), scalable system, with a current spin capacity reaching 96k, which is readily scalable to larger sizes[57]. To enable fair cross-platform comparisons, all experimental results were normalized to a standardized benchmark with 100 Ising spins and $10^4$ iterations (see Supplementary Note 10). Under this configuration, VSIM achieves the highest energy efficiency ($2.5\times10^4$ solutions per second per watt), representing an improvement of $1.5\times10^3$-$3.6\times10^7$ times over competing platforms. Moreover, VSIM demonstrates superior capability in solving sophisticated, real-world problems, such as global routing and layer assignment in EDA, highlighting its potential for industrial deployment.

| | | GPU[21] | QPU[13,14] | CIM[16,17] | PTNO[35] | SOT-MTJ[41] | SMTJ[31,39] | V-MTJ[43] | VSIM |
|---|---|---|---|---|---|---|---|---|---|
| Ising spin | Form | PRNG | Qubit | DOPO[Ψ] | PTNO | SOT-MTJ | SMTJ | V-MTJ | VCMA-MTJ |
| | Number | N/A[✝] | 5627 | 100k | 8 | 1 | 80 | 1143 | 96k |
| | Technology | CMOS chip | N/A | N/A | Devices | Devices | Devices | Devices | VC-MRAM chip |
| | Speed | 3.2 μs | 7 ns | 5 μs | 1.85 μs | 4 μs | 100 μs | 160 ns | < 1 ns |
| | Energy consumption | 38 μJ | -[✝] | - | 47 pJ | 810 pJ | 50 pJ | 4.7 pJ | < 40 fJ |
| System implementation | Connectivity | All-to-all | Sparse | All-to-all | All-to-all | All-to-all | All-to-all | All-to-all | All-to-all |
| | Temperature | 300 K | 12 mK | 300 K | 300 K | 300 K | 300 K | 300 K | 300 K |
| | Power | 450 W | 25 kW | - | 32 mW[‡] | - | 329 mW | 404 mW | 40 mW |
| Experimental application and performance | Type of COP | Max-cut | Max-cut | Max-cut | Max-cut | Integer-factorization | TSP | Integer-factorization | RSMT DVLA |
| | Difficulty level | ★ | ★ | ★ | ★ | ★★★ | ★★★ | ★★★ | ★★★★★ |
| | Solutions per second per watt[†] | $6.95\times10^{-4}$ | $5.71\times10^{-3}$ | - | $1.69\times10^{1}$ | - | $3.04\times10^{-2}$ | $1.55\times10^{1}$ | $2.50\times10^{4}$ |

**Table 1 | Comparison of VSIM with state-of-the-art Ising machines.** [Ψ] Degenerate optical parametric oscillator. [✝] " - " indicates that the data is not reported in the corresponding reference; "N/A" denotes not applicable or not suitable for direct comparison. [‡] An estimated extrapolation to a 100-spin system. [†] Normalized to a reference task involving a 100-spin system executing $10^4$ iterations to facilitate a fair comparison.

Fundamentally, QPUs offer rapid spin updates yet become highly time-consuming when tackling dense Ising topologies[14]. Their reliance on cryogenic cooling leads to extremely high power consumption (25 kW), resulting in energy efficiency seven orders of magnitude lower than that of VSIM. Similarly, optical approaches such as CIMs support large spin counts but require kilometer-scale fiber cavities[16,17], which severely impact overall system compactness and performance. In the electronic domain, GPUs suffer from inefficient Ising emulation and high power costs. Oscillator-based approaches (e.g., PTNO) are faced with restricted programmability and significant power and area overheads due to their reliance on passive coupling components (e.g., capacitors/resistors)[35]; furthermore, the simple bistable dynamics of such oscillators do not fully capture the stochastic nature of Ising spins, thereby compromising the system's convergence to high-quality solutions. In contrast, spintronic approaches offer highly tunable probabilistic behavior, enabling the system to effectively escape local minima. Crucially, VSIM surpasses other MTJ-based implementations (SOT-MTJ, SMTJ and V-MTJ) by two orders of magnitude improvement in both speed and energy consumption per spin update. Notably, the 160 ns spin-update latency observed in V-MTJ arises from two distinct factors. First, generating a 16-bit random number requires 16 sequential updates, each taking approximately 10 ns per stochastic bit[43]. Second, the intrinsic timescale of its magnetization dynamics is inherently long, as the magnetization has to reach a stable state that yields a 50% switching probability[58]. Our findings underscore that integrating a well-suited physical phenomenon, exemplified by pulse-width-tunable probabilistic switching of the VCMA-MTJ, with robust chip-level architectures capable of addressing large-scale problems, is key to unlocking substantial gains in both hardware performance and algorithm efficiency.

## Conclusions and outlook

In conclusion, we have demonstrated the first chip-scale spintronic Ising machine using VC-MRAM. Leveraging the tunable probabilistic behavior of VCMA-MTJs, we construct Ising spins that achieve single-spin update speeds below 1 ns and energy consumption under 40 fJ, both representing a 100× improvement over prior spintronic approaches. We fabricated an VC-MRAM chip integrating 96k Ising spins and developed VSIM, a scalable, ultra-fast, and ultra-low-power system for tackling NP-hard COPs, including global routing and layer assignment problems in the EDA domain. Experimental evaluations show that VSIM achieves an energy efficiency of $2.5\times10^4$ solutions per second per watt—three to seven orders of magnitude higher compared to quantum, optical, and other electronic counterparts—paving the way for broader deployment in real-world applications.

Further improvements could be achieved by developing a system-on-chip that integrates VC-MRAM with a matrix-vector multiplication (MVM) accelerator, enabling a more powerful VSIM. The energy efficiency of the MVM operation could be boosted through computing-in-memory MRAM[59,60], unleashing the full potential of an all-spintronic Ising chip for edge computing. Furthermore, multi-chip Ising architectures[22,24], designed for adaptive configurability, could be applied to various industrial-scale NP-hard problems, acting as high-performance Ising clusters in cloud computing. Collectively, these directions are expected to significantly improve the hardware efficiency and adaptability of physics-inspired computing systems, broadening their role in general-purpose computing for machine intelligence.

## Methods

### VC-MRAM fabrication

The VCMA-MTJs were fabricated on a 300-mm MRAM pilot line. As illustrated in Fig. 2a, the multilayer stack was deposited in situ at room temperature using physical vapor deposition in a 300-mm cluster tool (Canon-Anelva EC7800), followed by a post-deposition annealing process at 350°C for 1 hour under a 1 T magnetic field. The MgO tunnel barrier exhibited an RA of 300 Ω·μm$^2$. Circular MTJ devices with a nominal diameter of ~70 nm were patterned using 193-nm immersion lithography and sequentially etched via ion beam etching at both normal and grazing incidence angles. Electrical measurements revealed a room-temperature device resistance of ~80 kΩ and a magnetoresistance ratio of ~120%. Finally, a dual damascene copper top electrode was integrated using standard back-end-of-line processes to establish electrical interconnects.

The VC-MRAM chip integrates readout circuitry and high-speed write circuitry capable of generating pulse widths as short as 0.3 ns. A scan-based interface is used to configure both the pulse delay and pulse width for precise timing control during write operations. The CMOS-integrated dies were assembled in a 144-pin Low-Profile Quad Flat Package and mounted onto a chip holder for FPGA-based testing and VSIM implementation.

### Probability Measurement

The $P_{sw}$ versus magnetic field characteristics under varying voltage biases (Fig. 2c) were obtained using a custom-designed ultrafast probe card system. This system enables rapid, high-precision measurements of resistance-magnetic field (R-H) hysteresis loops. For each voltage bias condition, we performed 500 consecutive measurements to ensure statistical robustness. The VC-MRAM test system, incorporating a Xilinx PYNQ-Z2 FPGA development board and a custom-designed PCB hosting the integrated VC-MRAM chip, was used to measure the chip's probabilistic switching characteristics. The FPGA (part of the Zynq-7000 SoC on the PYNQ-Z2 board) orchestrated the measurement sequence: controlling W/R timing for the VC-MRAM chip, setting write voltages via separate on-PCB DACs (TLV5618A), triggering the VC-MRAM's on-chip pulse generator to apply voltage pulses of specific widths to the cells, and reading the resulting cell states through the Xilinx 7 series XADC module located on the PYNQ-Z2 board. Switching probabilities were determined by executing 1000 W/R cycles per cell for every tested pulse width setting.

### Implementation of VSIM

Extended Data Fig. 1 illustrates a closed-loop VSIM demonstration platform built with a VC-MRAM chip, a custom-designed PCB, and a Xilinx PYNQ-Z2 FPGA development board. The process begins in the PYNQ-Z2's processing system (PS), where an NP-hard COP is mapped to the Ising model to determine parameters $J$ and $h$. These parameters are then transferred to the PYNQ-Z2's programmable logic (PL) via the Advanced eXtensible Interface bus. The PL performs MVM operations to calculate the required pulse width and subsequently configures the VC-MRAM's pulse generator accordingly. After triggering a write pulse, the PL reads the resulting spin states from the VC-MRAM. Once the maximum iteration count is reached, the PL sends the final spin states back to the PS, and the corresponding COP solution is displayed on a screen.

### Evaluation of device variations

To investigate the impact of device variations, we utilized a full-FPGA implementation of the Ising machine for simulation. In contrast to the VSIM platform which employs VC-MRAM as Ising spins, this FPGA implementation emulates spin behavior using LFSRs, sigmoid LUTs, and digital comparators. Other operations such as MVM on the PL and the communication between the PS and PL were kept identical to those in the VSIM. To simulate devices with varying degrees of sub-100% PSP, the maximum output values of the sigmoid LUTs were configured from 50% to 90% in 10% increments. For each specific configuration of PSP and maximum iteration count, the full-FPGA Ising machine attempted to solve a target COP 100 times to determine the corresponding success probability.

## Acknowledgements

We acknowledge financial support from the Beijing Natural Science Foundation (QY24139), National Natural Science Foundation of China (62404015) and the Fundamental Research Funds for the Central Universities.

## Authors' contributions

W.Z. and S.L. initialized and supervised the project. S.L. and Y. Z. contributed equally to this work. S.L., Y. Z., A.L., Z.Z. and D.W.

fabricated the VC-MRAM chip. Y. Z. and S.L. performed the measurements and the implementation of VSIM with the help from A.L., L.Z., P.W. and L.G.. Y. Z., S.L., A.L. and W.Z. drafted the manuscript. All authors discussed the results and implications.

## Competing interests

The authors declare no competing interests.

## Additional information

1. Supplementary Information
2. Supplementary Videos
3. Supplementary Data

**Correspondence and requests for materials** should be addressed to Sai Li or Weisheng Zhao

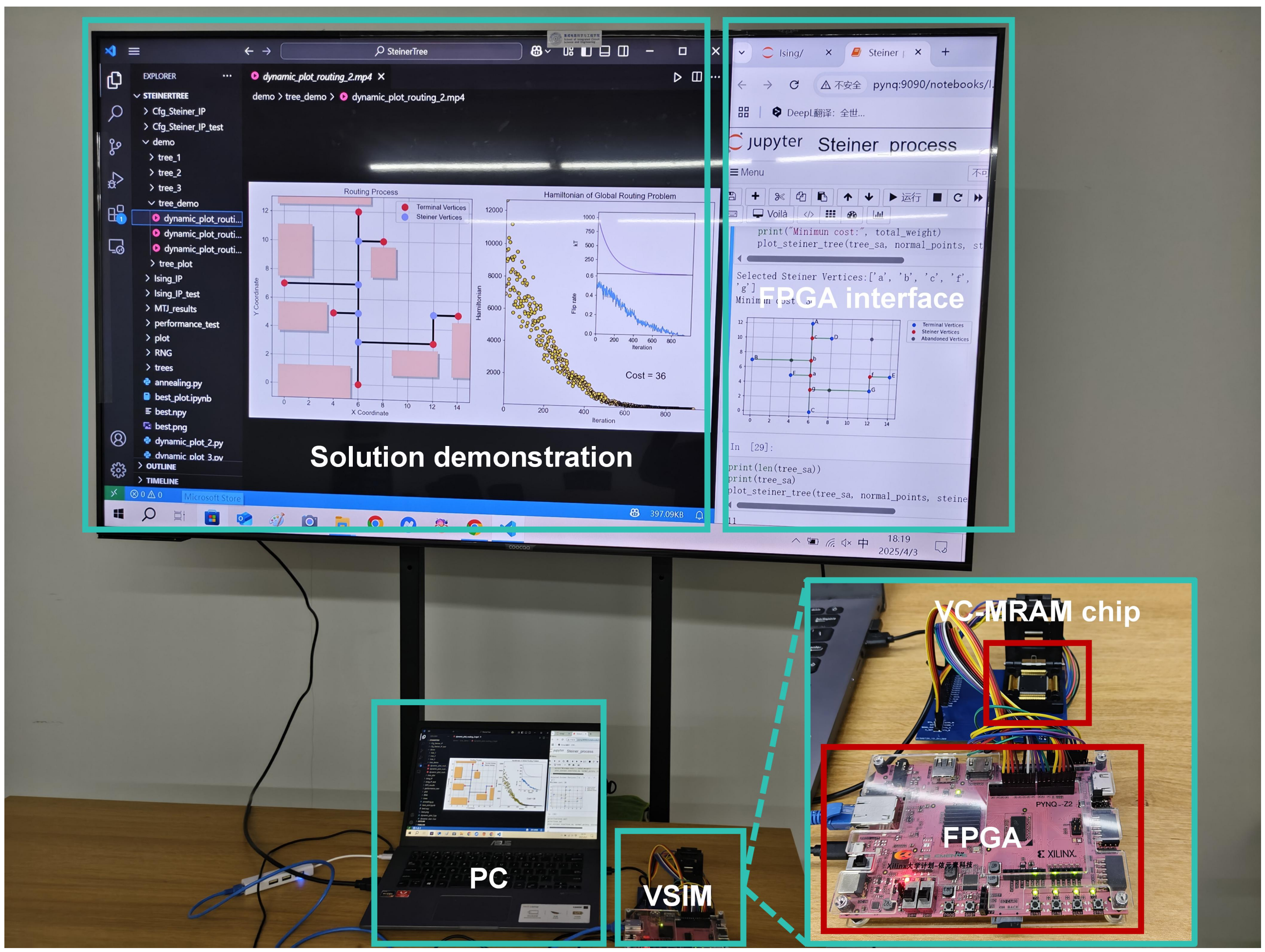

**Extended Data Fig. 1 | VSIM demonstration platform.** This system comprises a VC-MRAM chip mounted on a custom PCB interfaced with a Xilinx PYNQ-Z2 FPGA board, all controlled by a host PC. Operation begins on the PC, where a COP is mapped to an Ising model and parameters are configured using a Jupyter Notebook interface. These configurations are transferred to the FPGA, which then drives the VC-MRAM chip through iterative Ising computations. Finally, the resulting spin states, corresponding to the COP solution, are returned to the PC for visualization.

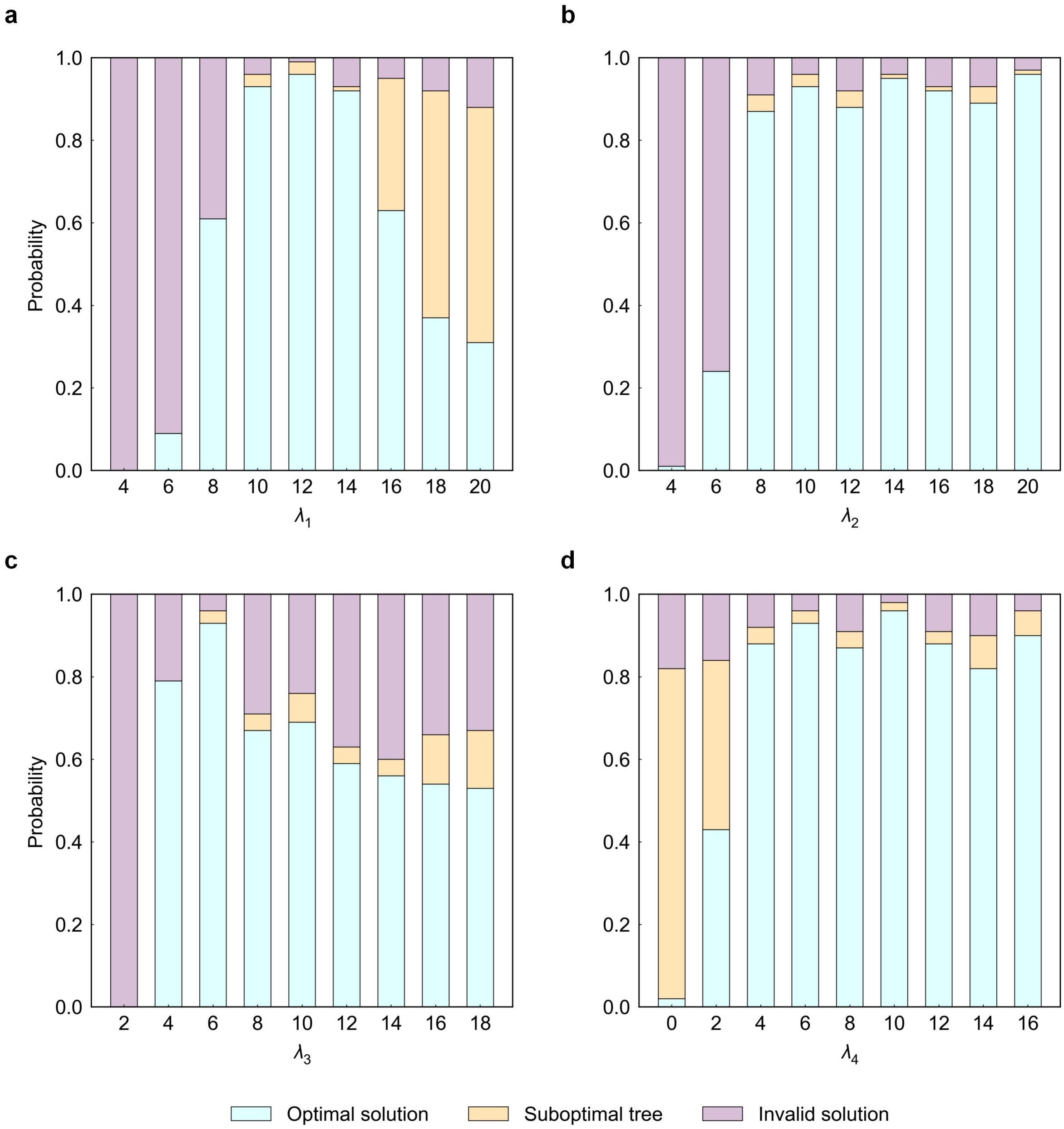

**Extended Data Fig. 2 | Influence of penalty coefficients on global routing solution probability.** Baseline coefficients are set to $\lambda_1 = \lambda_2 = 10$ and $\lambda_3 = \lambda_4 = 6$. Each panel (**a-d**) shows the probability variation when only one coefficient is changed. Effects of coefficient values: (i) Low $\lambda_1$, $\lambda_2$, and $\lambda_3$ lead to invalid solutions. (ii) Low $\lambda_4$ yields suboptimal trees, highlighting its importance for finding the ground state. (iii) High $\lambda_1$ or $\lambda_3$ reduce success probability due to conflicts among constraints or with the optimization objective. (iv) High $\lambda_2$ or $\lambda_4$ do not reduce success probability, implying no critical conflicts.

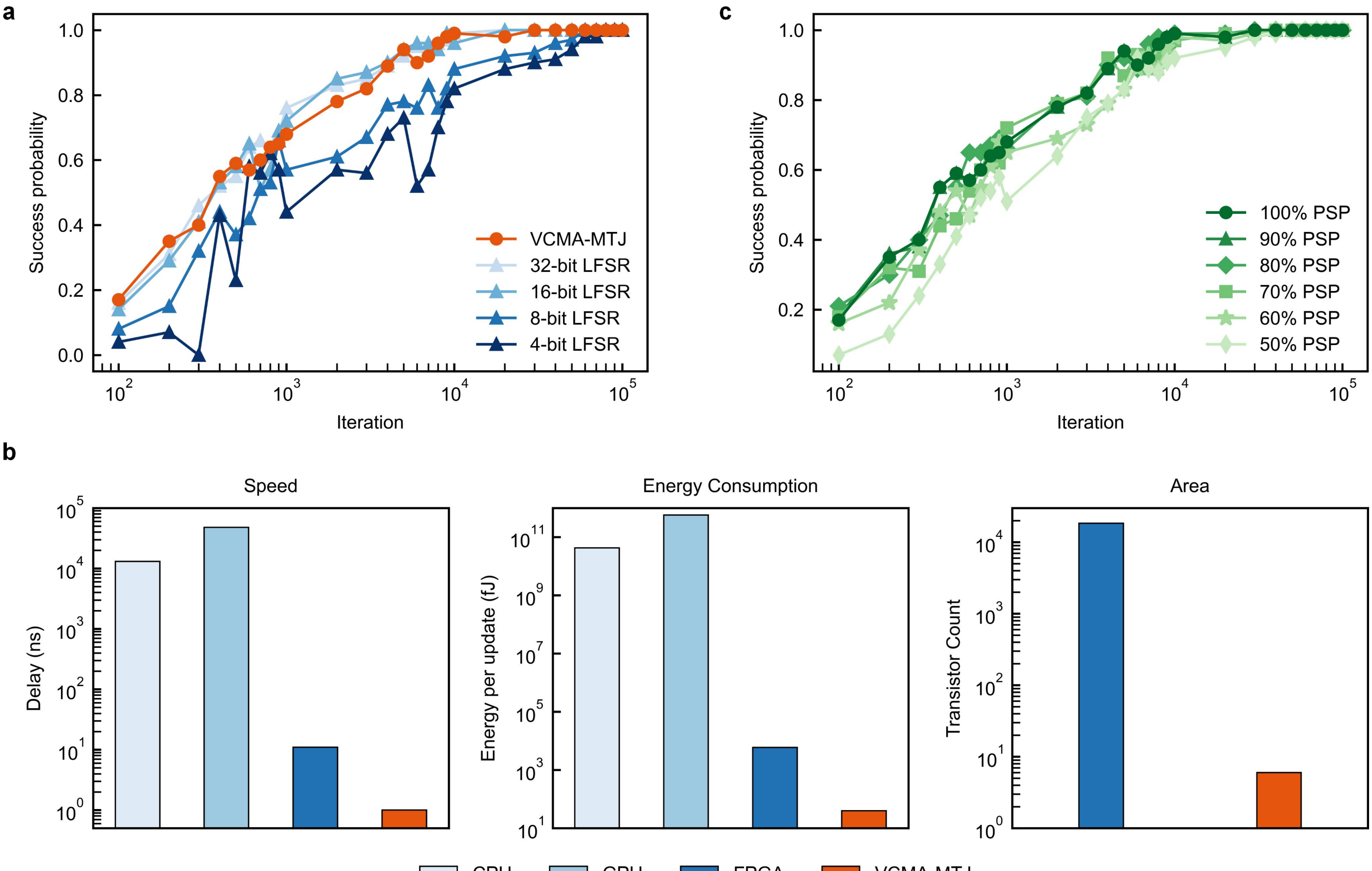

**Extended Data Fig. 3 | Performance evaluation of VCMA-MTJ as Ising spin for global routing applications. a,** Success probability versus iterations comparing the VCMA-MTJ and LFSRs with varying bit-widths. **b,** Hardware performance benchmarking (speed, area, energy consumption) of the VCMA-MTJ compared to CPU (Intel Xeon Platinum 8352V), GPU (NVIDIA GeForce RTX 4090) and FPGA (Zynq-7000) implementations. **c,** System robustness against device-to-device variations, demonstrating that the success probability remains largely unaffected when the PSP exceeds 60%.